\documentclass{elsart}
\usepackage{epsfig}
\usepackage{graphicx}
\usepackage{amssymb}
\journal{Nuclear Physics A}
\def\bbox#1{\mbox{\boldmath $#1$}}
\def\oo{$^{16}$O+$^{16}$O\ }
\def\ptx{$^{16}$O+$^{16}$O$^*$\ }
\def\2x{$^{16}$O+$^{16}$O$_{2^+}$\ }
\def\3x{$^{16}$O+$^{16}$O$_{3^-}$\ }
\def\o17o15{$^{16}$O($^{16}$O,$^{17}$O)$^{15}$O\ }
\def\InOO{$^{16}{\rm O}(^{16}{\rm O},^{16}$O$^{'})^{16}{\rm O}^*$\ }

\def\aA{$\alpha$-nucleus }
\def\AA{nucleus-nucleus\ }
\begin{document}
\begin{frontmatter}
\title{Study of refractive structure in the inelastic $^{\bf 16}$O+$^{\bf 16}$O
scattering at the incident energies of 250 to 1120 MeV\thanksref{Ackn}}
\thanks[Ackn]{Research supported, in part, by the
Alexander-von-Humboldt Stiftung of Germany, Natural Science Council of Vietnam
and Vietnam Atomic Energy Commission (VAEC).}
\author[INST,HMI]{Dao T. Khoa\corauthref{cor1}},
\ead{khoa@vaec.gov.vn} \corauth[cor1]{Corresponding author.}
\author[HMI]{H.G. Bohlen}, \ead{bohlen@hmi.de}
\author[HMI,FU]{W. von Oertzen}, \ead{oertzen@hmi.de}
\author[TUE]{G. Bartnitzky},
\author[HMI,TUE]{A. Blazevic},
\author[TUE]{F. Nuoffer},
\author[HMI]{B. Gebauer},
\author[GAN]{W. Mittig} and
\author[GAN]{P. Roussel-Chomaz}
\address[INST]{Institute for Nuclear Science and
 Technique, VAEC, P.O. Box 5T-160,\\ Nghia Do, Hanoi, Vietnam.}
\address[HMI]{Hahn-Meitner-Institut GmbH, Glienicker Str. 100, D-14109 Berlin,
 Germany.}
\address[FU]{Fachbereich Physik, Freie Universit\"at Berlin.}
\address[TUE]{Physikalisches Institut, Universit\"at T\"ubingen,
  Auf der Morgenstelle 14, D-72076 T\"ubingen, Germany.}
\address[GAN]{GANIL, Bd. Henri Becquerel, BP 5027, F-14021 Caen Cedex,
France.}

\begin{abstract}
The data of inelastic \oo scattering to the lowest 2$^+$ and 3$^-$ excited
states of $^{16}$O have been measured at $E_{\rm lab}=250, 350, 480, 704$ and
1120 MeV and analyzed consistently in the distorted wave Born approximation
(DWBA), using the semi-microscopic optical potentials and inelastic form factors
given by the folding model, to reveal possible refractive structure of the
nuclear rainbow that was identified earlier in the elastic \oo scattering
channel at the same energies. Given the known transition strengths of the
2$^+_1$ and 3$^-_1$ states of $^{16}$O well determined from the $(e,e')$ data,
the DWBA description of the inelastic data over the whole angular range was
possible only if the absorption in the exit channels is significantly increased
(especially, for the $^{16}$O+$^{16}$O$_{2^+}^*$ exit channel). Although the
refractive pattern of the inelastic \oo scattering was found to be less
pronounced compared to that observed in the elastic scattering channel, a clear
remnant of the main rainbow maximum could still be seen in the inelastic cross
section at $E_{\rm lab}=350-704$ MeV.
\end{abstract}

\end{frontmatter}

\section{Introduction}
The study of elastic, refractive scattering of light \AA systems, like \oo and
$^{16}$O,$^{12}$C+$^{12}$C, has proven to be very helpful for our understanding
of the interaction between heavy ions (HI) \cite{Bra97}. The main reason is that
the refractive structures or \emph{nuclear rainbow} observed in the \AA elastic
scattering can provide information on the HI optical potential at small
inter-nuclear distances. Although most of the studies have been concentrated on
the elastic scattering (see, e.g., Ref.~\cite{Kho00} for the \oo system). The
refractive effects have also been found and discussed for one-neutron transfer
reactions in the $^{12}$C+$^{12}$C, $^{13}$C+$^{12}$C \cite{Boh85,Sat89} and \oo
systems \cite{Boh02}. For example, in the \oo system, where the elastic
scattering has shown a pronounced nuclear rainbow pattern \cite{Kho00}, a clear
remnant of the rainbow pattern was identified in the \o17o15 transfer cross
section for the ground state transition to $^{15}$O$_{{1/2}^-}$ \cite{Boh02}.
Beside the transfer channel, the inelastic scattering channel of a refractive
\AA system can also have a well defined rainbow pattern \cite{Kho87}. In
particular, it has been shown recently by Michel and Ohkubo \cite{Mic04} that
the transparency seen in some light HI in the form of Airy structure in the
elastic scattering channel could have similar pattern in the inelastic
scattering channel.

To complete our study \cite{Kho00,Boh02} of the refractive structure in the
quasi-elastic \oo scattering, we present in this work the results of our
analysis in the distorted wave Born approximation (DWBA) of the inelastic \InOO
scattering data which have been measured in parallel with the elastic \oo
scattering at $E_{\rm lab}=250, 350, 480, 704$ and 1120 MeV
\cite{Boh93,Sti89,Bar96,Nuo98}. These data were taken for the transitions to the
lowest 2$^+$ and $3^-$ states in $^{16}$O at 6.92 MeV and 6.13 MeV excitation
energy, respectively, and cover about the same wide angular range as that
covered by the elastic scattering and one-neutron transfer data. Since the
transition strengths of the 2$^+_1$ and 3$^-_1$ states of $^{16}$O have been
accurately determined from the $(e,e')$ data \cite{Ram01,Kib02}, they can serve
as constraints for the deformation lengths used to generate the inelastic form
factors for our DWBA analysis. As a consequence, the measured inelastic
scattering data provide a good database for the study of refractive features in
the inelastic \InOO scattering.

A brief survey of the measurements of the inelastic \oo scattering to the
2$^+_1$ and 3$^-_1$ states of $^{16}$O is presented in Sect.~\ref{sec1}. The
basic ingredients of the semi-microscopic folding approach to calculate the
optical potential and inelastic form factors for the optical model (OM) and DWBA
analyses are given in Sect.~\ref{sec2}. The DWBA results of the analysis of the
inelastic \InOO scattering data at $E_{\rm lab}=250-1120$ MeV are discussed in
Sect.~\ref{sec3}. A summary and the main conclusions of this work are given in
Sect.~\ref{sec4}.

\section{Inelastic $^{16}{\rm O}(^{16}{\rm O},^{16}{\rm O}^{'})^{16}{\rm O}^*$
 data}
 \label{sec1}
The experimental spectra have been taken at the incident energies $E_{\rm
lab}=250-480$ MeV \cite{Boh93,Sti89,Bar96} using the Q3D magnetic spectrograph
at HMI, and at 704 and 1120 MeV \cite{Bar96,Nuo98} using the SPEG spectrograph
at GANIL. The spectral lines of the 3$^-$ state at 6.13~MeV and 2$^+$ state at
6.92~MeV were observed in these spectra as the strongest inelastic transitions.
The method used to obtain the experimental cross sections of inelastic
scattering from the analysis of the inelastic spectra is described in the
following.

Since the projectile and target are identical particles and the same state can
be excited in the projectile as well as in the target (and also in both by
mutual excitation), every excited state of $^{16}{\rm O}$ appears in the
spectrum \emph{twice} at the same position of the corresponding excitation
energy: (1) a sharp line is observed for the target excitation, and (2) a broad
line for the projectile excitation (the mutual excitation could not be inferred
from our data due to a small cross section and an increasing three-body
background). The line-broadening observed for the excited states of the
projectile is due to the recoil momentum of the photon emitted in-flight by the
projectile. This effect is directly related to the Doppler-broadening usually
observed in the $\gamma$-ray spectroscopy. The maximum energy shift in the
laboratory system is given by $\Delta E_{\rm max}= \pm (v/c)E_{\gamma }$
\cite{Boh76}, where $E_{\gamma }$ is the energy of the $\gamma$-transition and
$v$ is the velocity of the photon-emitting outgoing $^{16}{\rm O}$ projectile.
These maximum shifts can be rather large: at the forward scattering angles
they are, e.g., for the $\gamma$-transition from the 2$^+$ state at 6.92~MeV to
the ground state $\Delta E_{\rm max}=\pm 1.45$~MeV at the beam energy $E_{\rm
lab} = 350$~MeV, and $\Delta E_{\rm max}=\pm 2.45$~MeV at $E_{\rm lab}=1120$~MeV
(angular shifts, which also result from the $\gamma$-recoil, are small and not
larger than the angular resolution). Note that at 1120 MeV, the base width of
the broadened line is almost 5~MeV wide, and this needs to be compared with the
experimental energy resolution of about 0.5~MeV at both incident energies.

\begin{figure}[t] \hspace*{-0.7cm}
\includegraphics[angle=0,scale=1.5]{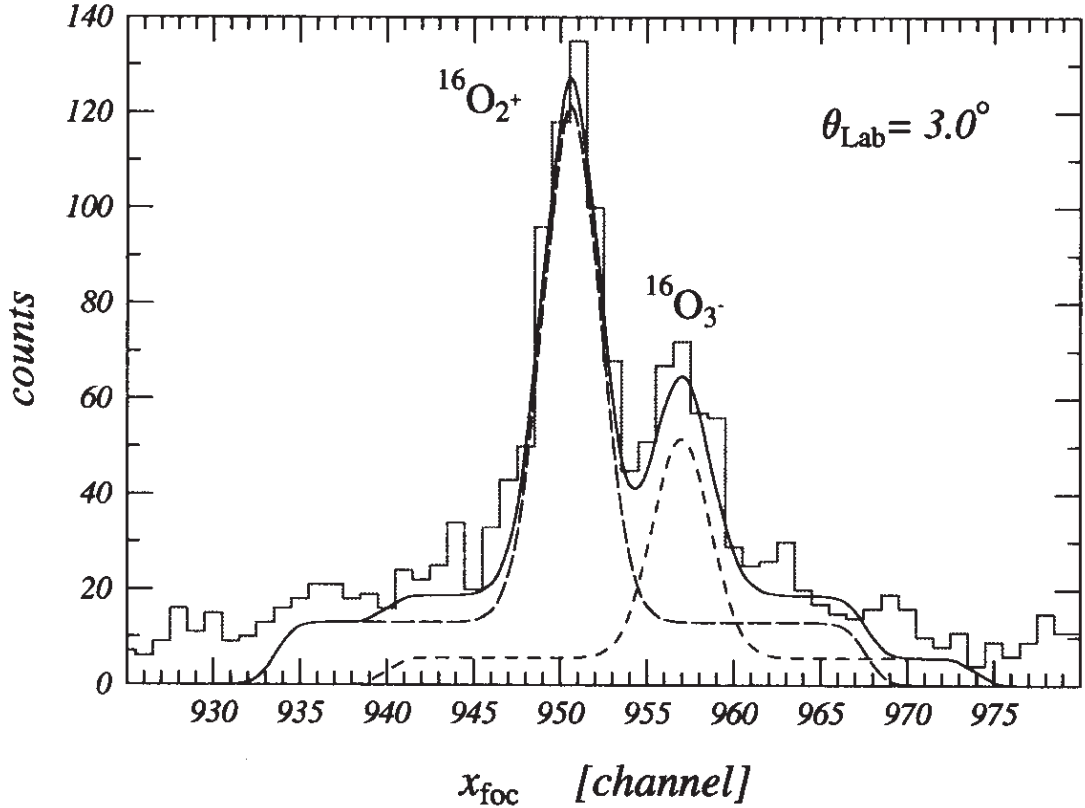}
\caption{Spectrum of the inelastic \InOO scattering at $E_{\rm lab}=1120$~MeV
and $\Theta_{\rm lab}=3.0^\circ$ in the excitation energy region of the 3$^-$
and 2$^+$ excited states at 6.13 MeV and 6.92 MeV, respectively. The line shape
used for the fit is described in the text.} \label{f0} \vspace{0.5cm}
\end{figure}

In the spectrum of inelastic excitation the lines resulting from the target
excitation are described by Gaussians using the experimental resolution as the
width, and the projectile excitation by rectangular distributions with base
widths $2(v/c)E_{\gamma}$ which were folded with the experimental resolution.
For a given excited state the number of counts for the projectile and target
excitation must be equal. This is required by the symmetry of the system which
implies the \emph{equal} excitation strengths of the same excited state in both
the projectile and target nuclei at the center-of-mass (c.m.) scattering angles
$\Theta_{\rm c.m.}$ and $\pi-\Theta_{\rm c.m.}$. Thus, the Gaussian line of the
target excitation and the broad line of the projectile excitation have been
first normalized to the same area before they were added. The resulting
distributions were then fitted to the corresponding excited states in the
spectrum by optimizing the normalizations. The fit can be performed
simultaneously for several different excited states of $^{16}{\rm O}$ with
overlapping widths. Fig.~\ref{f0} shows a spectrum of the inelastic scattering
to the 3$^-$ and 2$^+$ excited states at 6.13~MeV and 6.92~MeV at $E_{\rm
lab}=1120$~MeV and $\Theta_{\rm lab}=3.0^\circ$. The number of counts obtained
in the fit of an excited state represents the sum of inelastic cross sections
for the projectile and target excitations, which was further divided by a factor
of two to obtain the single excitation cross section for the DWBA analysis.

In the analysis of the measured spectra we have used an experimental technique,
where two-dimensional plots of scattering angle versus position in the focal
plane are constructed to perform projections on either axis by applying specific
gate conditions. To obtain spectra like that shown in Fig.~\ref{f0} it is
necessary to set rather narrow angular gates (typically $0.2^\circ$ at small
angles and slightly larger at intermediate angles) before the spectra are
projected to the position axis. This procedure has been applied at the energies
of 250, 350, 704, and 1120~MeV up to the scattering angles of $22^\circ,
16.2^\circ, 7.2^\circ$, and $9^\circ$, respectively, in the laboratory frame. As
a result, the angular distributions of the 3$^-$ state at 6.13~MeV and 2$^+$
state at 6.92~MeV have been well separated in these angular ranges. At larger
angles it was not possible to separate the 3$^-$ and 2$^+$ excited states using
this method. Their sum could be, however, easily projected when the gate in
excitation energy was set in such a way that both broad lines were included
within the acceptable energy limits (in the energy range where other excited
states of $^{16}{\rm O}$ were not observed). In this way, the angular
distribution of the total (2$^++3^-$) cross section has been projected for these
states up to $42.5^\circ, 37^\circ$, and $19.1^\circ$ in the laboratory frame at
the energies of 250, 350, and 704~MeV, respectively. At 1120~MeV, the range
could not be extended due to a steep decrease of the cross sections. At 480~MeV,
only the angular distribution of the sum of the (2$^++3^-$) cross sections has
been projected.

We also note that the observed spectrum represents overwhelmingly the inelastic
scattering to the 3$^-$ and 2$^+$ excited states at 6.13~MeV and 6.92~MeV, and
the strength mixing by the 0$^+$ and 1$^-$ states at 6.05~MeV and 7.12~MeV can
be neglected. While no peak has been observed for the 0$^+$ state in the angular
bin covered by our measurement, a slight rise caused by the 1$^-$ state at
7.12~MeV could be seen. The dipole excitation strength of the 1$^-$ state is
however too weak compared to those of the 3$^-$ and 2$^+$ states:
$B(E1\downarrow)\approx 3.4\times 10^{-4}$ W.u. compared to
$B(E3\downarrow)\approx 13$ W.u. and $B(E2\downarrow)\approx 3.5$ W.u.
\cite{End93}, to make a sizable contribution to the total inelastic \InOO
scattering cross section.

\section{Main features of the Folding + DWBA analysis}
 \label{sec2}
The most important ingredients of the DWBA analysis of the inelastic scattering
\InOO are the \emph{complex} optical potentials (OP) in the entrance and exit
channels
\begin{equation}
U^{(0)}_{\rm in}=<^{16}{\rm O},^{16}{\rm O}|V|^{16}{\rm O},^{16}{\rm O}>,\
U^{(0)}_{\rm out}=<^{16}{\rm O}^*,^{16}{\rm O}|V|^{16}{\rm O}^*,
 ^{16}{\rm O}>, \label{e1}
\end{equation}
and the transition potential or inelastic form factor (FF)
\begin{equation}
 U^{(\lambda)}_{\rm trans}=
 <^{16}{\rm O}^*,^{16}{\rm O}|V|^{16}{\rm O},^{16}{\rm O}>,\
 {\rm with}\ \lambda=2,3. \label{e2}
\end{equation}
For the elastic and inelastic HI scattering, there are two main prescriptions of
constructing the nuclear matrix elements (\ref{e1}) and (\ref{e2}):

(I) The phenomenological approach of parameterizing the OP (\ref{e1}) in terms
of a conventional Woods-Saxon potential whose parameters to be adjusted to the
best fit of elastic scattering data. The inelastic FF (\ref{e2}) is then given
by the radial derivative of the OP, scaled by a factor known as the deformation
length. This prescription for the inelastic FF is known as the
deformed-optical-potential (DOP) model where the deformation length is normally
obtained by matching the calculated cross section to the observed inelastic data
\cite{Sat83}.

(II) The (semi-microscopic) folding approach \cite{Sat79} where the real OP and
inelastic FF are calculated microscopically using an appropriate effective
in-medium nucleon-nucleon (NN) interaction and the ground-state and transition
nuclear densities for the two colliding nuclei.

While the shape of the inelastic folded potential has a strong dependence on the
multipolarity $\lambda$ of the transition, the inelastic FF given by the simple
DOP model (I) has a $\lambda$-independent shape \cite{Be95,Be96}. As a result,
the nuclear deformation lengths extracted by the DWBA analysis using the DOP
model can be significantly underestimated \cite{Be95}, especially for high
multipoles $\lambda\geq 3$.

In the present work, following the success of the folding potential in the OM
description of the elastic \oo scattering \cite{Kho00} and DWBA analysis of the
\o17o15 one-neutron transfer data at the same energies \cite{Boh02}, we use a
recent version of the double-folding model \cite{KhoSat} to consistently
calculate the real OP and inelastic FF for the DWBA analysis of the inelastic
\InOO scattering. Within this double-folding approach \cite{KhoSat} the
projectile-target interaction potential in the elastic or inelastic scattering
channel is evaluated as an energy dependent Hartree-Fock-type potential of the
dinuclear system
\begin {equation}
 V_{\rm F}=\sum_{i,i'\in A_1;j,j'\in A_2}
 [<i'j'|v_{\rm D}|ij> +<i'j'|v_{\rm EX}|ji>]=V_{\rm D}+V_{\rm EX},
 \label{e3}
\end{equation}
where $v_{\rm D}$ and $v_{\rm EX}$ are the direct and exchange parts of the
effective NN interaction. The calculation of $V_{\rm D(EX)}$ is done iteratively
based on a density-matrix expansion method \cite{KhoSat,Kho01}. In this
calculation, the inputs for mass numbers and incident energies were taken as
given by the relativistically corrected kinematics \cite{Far84}. Like in our
earlier works \cite{Kho00,Boh02}, we have used for the effective NN interaction
$v_{\rm D(EX)}$ the CDM3Y6 version \cite{Kho97} of the energy- and density
dependent M3Y-Paris interaction. Since the density-dependent parameters of the
CDM3Y6 interaction were adjusted to reproduce correctly saturation properties of
the \emph{infinite} nuclear matter, to use this interaction for a system of two
\emph{finite} nuclei one needs to take into account the kinematical
transformation from the NN frame to the \AA frame. Thus, the CDM3Y6 interaction
was further scaled by a kinematics modification factor deduced from Eq.~(19) of
Ref.~\cite{Lov81}. At the incident energies considered here, this factor leads
only to a marginal change of the potential strength (from around 1\% at 250 MeV
to 5\% at 1120 MeV) compared to that given by the same folding approach
(\ref{e3}) but neglecting the kinematics modification factor.

To calculate consistently both the optical potential and inelastic form factor
one needs to take into account explicitly the multipole decomposition of the
nuclear density distribution $\rho(\bbox{r})$ that enters the folding
calculation (see details in Ref.~\cite{KhoSat})
\begin {equation}
 \rho_{JM\to J'M'}(\bbox{r})=\sum_{\lambda\mu}
 <JM\lambda\mu|J'M'>C_\lambda\rho_\lambda(r)
 [i^{\lambda}Y_{\lambda\mu}(\hat{\bbox{r}})]^*,
\label{e7}
\end{equation}
where $JM$ and $J'M'$ are the nuclear spins and their projections in the initial
and final states, respectively, $C_0=\sqrt{4\pi}$ and $C_{\lambda}$=1 for
$\lambda\neq 0$; $\rho_\lambda(r)$ is the nuclear transition density of the
$2^{\lambda}$-pole excitation. In the present work, we adopt the
collective-model Bohr-Mottelson prescription \cite{Bor75} to construct the
nuclear transition densities for the 2$^+$ and 3$^-$ excitations of $^{16}$O as
\begin {equation}
\rho_\lambda(r)=-\delta_\lambda\frac{d\rho_0(r)}{dr},\ {\rm with}\ \lambda=2,3.
\label{e8}
\end{equation}
Here $\rho_0(r)$ is the total ground state (g.s.) density and $\delta_\lambda$
is the deformation length of the 2$^+$ or 3$^-$ excitation of $^{16}$O. The g.s.
density of $^{16}$O was taken as a Fermi distribution with parameters
\cite{Far85} chosen to reproduce the shell-model density for $^{16}$O, and the
neutron and proton parts of the $^{16}$O density were assumed to have the same
shape. Such an \emph{isoscalar} assumption for the g.s. density implies
\cite{KhoSat} that the same nuclear deformation length is used for neutron and
proton parts of the transition density (\ref{e8}). These deformation lengths
were determined (see Eqs.~(3.33)-(3.35) in Ref.~\cite{KhoSat}) from the measured
electric transition strengths of the 2$^+_1$ and 3$^-_1$ states of $^{16}$O,
$B(E2\uparrow)=40.6\pm 3.8\ e^2$fm$^4$ \cite{Ram01} and $B(E3\uparrow)=1480\pm
50\ e^2$fm$^6$ \cite{Kib02}, so that $\delta_2=1.038\pm 0.048$ and
$\delta_3=1.825\pm 0.031$ fm were used to generate the transition densities
(\ref{e8}) for the 2$^+_1$ and 3$^-_1$ states. Given these inputs for the
nuclear densities, the (\emph{energy dependent}) real folded optical potential
$V^{(0)}_{\rm F}$ and transition form factors $V^{(2)}_{\rm F}$ and
$V^{(3)}_{\rm F}$ were obtained straightforwardly in the double-folding approach
\cite{KhoSat}. Along with the nuclear FF, the Coulomb inelastic form factor
$V^{(\lambda)}_{\rm C}$ was calculated by the same folding method (\ref{e3}),
using the g.s. and transition \emph{charge} densities of the two colliding
$^{16}$O nuclei and the Coulomb interaction between the two protons.

To have the \emph{complex} energy dependent optical potentials for the entrance
and exit channels, the real (folded) OP was further added, as in our previous
folding analyses, by a Woods-Saxon (WS) imaginary potential that contains both
the volume and surface terms. Thus, the OP used in the present DWBA analysis is
\begin{equation}
 U^{(0)}_{\rm in(out)}(R)=V^{(0)}_{\rm C}(R)+N^{(0)}_{\rm V}V^{(0)}_{\rm F}(R)
 -i[W_{\rm V}(R)+W_{\rm D}(R)]. \label{e9}
\end{equation}
Here $V^{(0)}_{\rm C}$ is the elastic Coulomb potential between a point charge
and a uniform charge distribution of radius $R_{\rm
C}=1.3\times(A_1^{1/3}+A_2^{1/3})$ fm and the volume and surface parts of the
imaginary WS potential are determined as
\begin{equation}
 W_{\rm V}(R)=W_{\rm V}{\{1+\exp[(r-R_{\rm V})/a_{\rm V}]\}}^{-1},
\label{e9v}
\end{equation}
\begin{equation}
 W_{\rm D}(R)=4W_{\rm D}\exp[(R-R_{\rm D})/a_{\rm D}]
 \{1+\exp[(R-R_{\rm D})/a_{\rm D}]\}^{-2}. \label{e9d}
\end{equation}
For $U^{(0)}_{\rm in}$, the renormalization factor $N^{(0)}_{\rm V}$ of the real
folded potential $V^{(0)}_{\rm F}$ and parameters of the WS imaginary potential
were adjusted in each case to the best fit of the measured elastic \oo
scattering data \cite{Kho00}. It is commonly assumed in the DWBA analysis of
inelastic \AA scattering \cite{Sat83}, especially when one uses the DOP model
for the inelastic FF, that $U^{(0)}_{\rm out}=U^{(0)}_{\rm in}$. In the present
work, given the highly accurate inelastic \InOO scattering data and inelastic
folded FF's fixed by the known transition strengths of the 2$^+_1$ and 3$^-_1$
states of $^{16}$O, we find that this assumption is not always a good
approximation, and it could lead to an underestimation of the transition
strength of the nuclear excitation when the refractive effect is strong.

Within the folding model \cite{KhoSat}, the following \emph{complex} energy
dependent inelastic FF is often used for the transition potential (\ref{e2})
\begin{equation}
 U^{(\lambda)}_{\rm trans}(R)=V^{(\lambda)}_{\rm C}(R)+
 N^{(\lambda)}_{\rm V}V^{(\lambda)}_{\rm F}(R)-
 i\delta_\lambda\frac{dW_{\rm in}(R)}{dR},\ {\rm where}\ \lambda=2,3, \label{e10}
\end{equation}
and $W_{\rm in}(R)=W_{\rm V}(R)+W_{\rm D}(R)$ is the WS imaginary part of
$U^{(0)}_{\rm in}(R)$. With the deformation length fixed by the known transition
strength, the renormalization factor $N^{(\lambda)}_{\rm V}$ of the real folded
FF is usually adjusted to the best DWBA fit of inelastic scattering data and it
should be close to unity for the folding procedure (\ref{e10}) to be meaningful.
All the OM and DWBA calculations were made using the code ECIS97 written by
Raynal \cite{Ray97}.

\section{Results of the DWBA analysis and discussion}
 \label{sec3}
To have a reliable DWBA description of the inelastic \InOO scattering, one needs
first to determine the parameters of the OP (\ref{e1}) for the entrance and exit
channels. In addition to the scaling the CDM3Y6 interaction by a kinematics
modification factor, a more accurate local approximation \cite{Kho01} was also
used for the exchange potential $V_{\rm EX}$ compared to the folded potentials
used earlier in Ref.~\cite{Kho00}. Therefore, we have made again a detailed OM
analysis of the elastic \oo scattering data \cite{Boh93,Sti89,Bar96,Nuo98} using
the complex OP (\ref{e9}). The obtained OP parameter sets (see the best-fit OP
parameters for the \oo entrance channel at different incident energies in
Table~\ref{t1}) reproduce the same refractive structure in the elastic \oo cross
sections as that reported in Ref.~\cite{Kho00}. Note that the improved folding
procedure leads to a slightly smaller renormalization factor $N^{(0)}_{\rm V}$
of the real folded potential at all energies, with the largest difference of
about 10\% found at 250 MeV. The corresponding volume integrals $J_{\rm V}$ are
changed only by about 2-5\% and they indicate practically the same potential
family for the real OP as that discussed in Ref.~\cite{Kho00}.
\begin{table}\small
\caption{\small OP parameters (\ref{e9})-(\ref{e9d}) for the entrance and exit
channels used in the DWBA analysis of the inelastic \InOO scattering at incident
energies of 250 to 1120 MeV. $J_{\rm V}$ and $J_{\rm W}$ are the volume
integrals (per interacting nucleon pair) of the real folded potential and of the
imaginary WS potential, respectively. If blank, the values are the same as those
in the preceding row.} \label{t1}
\begin{tabular}{c l c c c c c c c c c} \hline\small
 $E_{\rm lab}$ & channel & $N^{(0)}_{\rm V}$ & $J_{\rm V}$ & $W_{\rm V}$ &
 $R_{\rm V}$ & $a_{\rm V}$ & $W_{\rm D}$ & $R_{\rm D}$ & $a_{\rm D}$ &
 $J_{\rm W}$ \\
(MeV) & & & (MeVfm$^3$) & (MeV) & (fm) & (fm) & (MeV) & (fm) & (fm) &
(MeVfm$^3$) \\ \hline
 250 & entrance & 0.781 & 290.0 & 38.34 & 4.405 & 0.949 & 6.630 &
 5.657 & 0.483 & 98.81 \\
 & exit (2$^+$) &  &  & 48.46 &  &  &  &  &  & 119.4 \\
 & exit (3$^-$) &  &  & 34.87 &  &  &  &  &  & 91.76 \\ \hline
350 & entrance & 0.845 & 298.2 & 25.32 & 5.832 & 0.629 & 6.932 &
 4.846 & 0.350 & 103.0 \\
 & exit (2$^+$) &  &  & 35.76 &  &  &  &  &  & 140.8 \\
 & exit (3$^-$) &  &  & 27.00 &  &  &  &  &  & 109.1 \\ \hline
480 & entrance & 0.791 & 261.5 & 31.59 & 5.421 & 0.711 & 1.647 &
 5.114 & 0.415 & 99.88 \\
 & exit (2$^+$) &  &  & 41.89 &  &  & 15.75 &  &  & 161.9 \\
 & exit (3$^-$) &  &  & 33.89 &  &  & 6.586 &  &  & 117.6 \\ \hline
704 & entrance & 0.867 & 256.1 & 28.62 & 5.669 & 0.652 & 7.095 &
 4.153 & 0.450 & 107.7 \\
 & exit (2$^+$) &  &  & 39.96 &  &  & 6.010 &  &  & 144.2 \\
 & exit (3$^-$) &  &  & 32.14 &  &  & 6.862 &  &  & 119.1 \\ \hline
1120 & entrance & 0.905 & 218.6 & 46.34 & 4.793 & 0.809 & 0.868 &
 5.640 & 0.391 & 109.1 \\
 & exit (2$^+$) &  &  & 61.98 &  &  & 3.888 &  &  & 152.7 \\
 & exit (3$^-$) &  &  & 50.30 &  &  & 1.178 &  &  & 119.0 \\ \hline
\end{tabular}
\end{table}

\begin{figure}[t] \hspace*{-1cm}
\includegraphics[angle=0,scale=0.81]{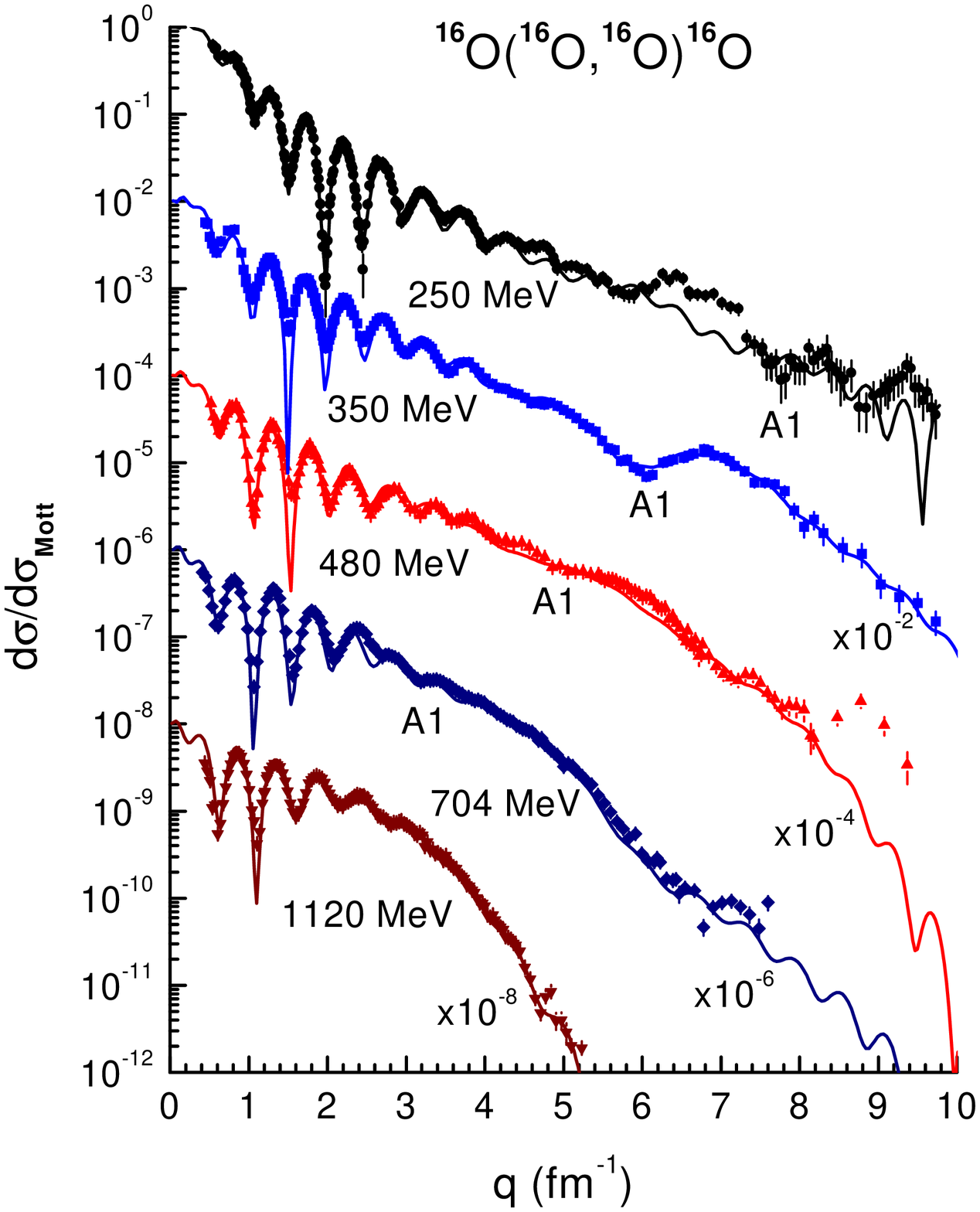}\vspace{-1.5cm}
\caption{The OM description of the elastic \oo scattering data measured at
$E_{\rm lab}=250-1120$ MeV \cite{Boh93,Sti89,Bar96,Nuo98} given by the
semi-microscopic optical potential (\ref{e9}). The OP parameters are those
determined for the entrance \oo channel (see Table~\ref{t1}). The cross sections
are given in ratio to the Mott cross section and plotted versus the momentum
transfer $q=2k\sin(\Theta_{\rm c.m.}/2)$, where $k$ is the wave number of the
projectile. A1 indicates location of the first Airy minimum.} \label{f1}
\end{figure}

To show more clearly the evolution of the refractive Airy pattern with the
increasing incident energy, we have plotted in Fig.~\ref{f1} the measured
elastic \oo data and the calculated cross sections versus the momentum transfer
$q=2k\sin(\Theta_{\rm c.m.}/2)$, where $k$ is the wave number of the projectile.
In such a presentation, the location of the first Airy minima (A1) can be traced
up to the energy of 704 MeV. The rainbow pattern associated with A1 \cite{Kho00}
is somewhat obscured at $E_{\rm lab}=250$ MeV by the Mott interference caused by
boson symmetry between the two identical $^{16}$O nuclei, but it becomes more
prominent at $E_{\rm lab}=350$ MeV. Actually, the elastic data at 350 MeV
\cite{Sti89} is now known as a very clear experimental evidence of the nuclear
rainbow scattering pattern in elastic heavy-ion scattering. In the present work
we investigate whether this rainbow pattern also exists in the inelastic \InOO
scattering channels.

\begin{figure}[htb]
\hspace*{-1.5cm} \vspace{-2cm}
\includegraphics[angle=0,scale=0.81]{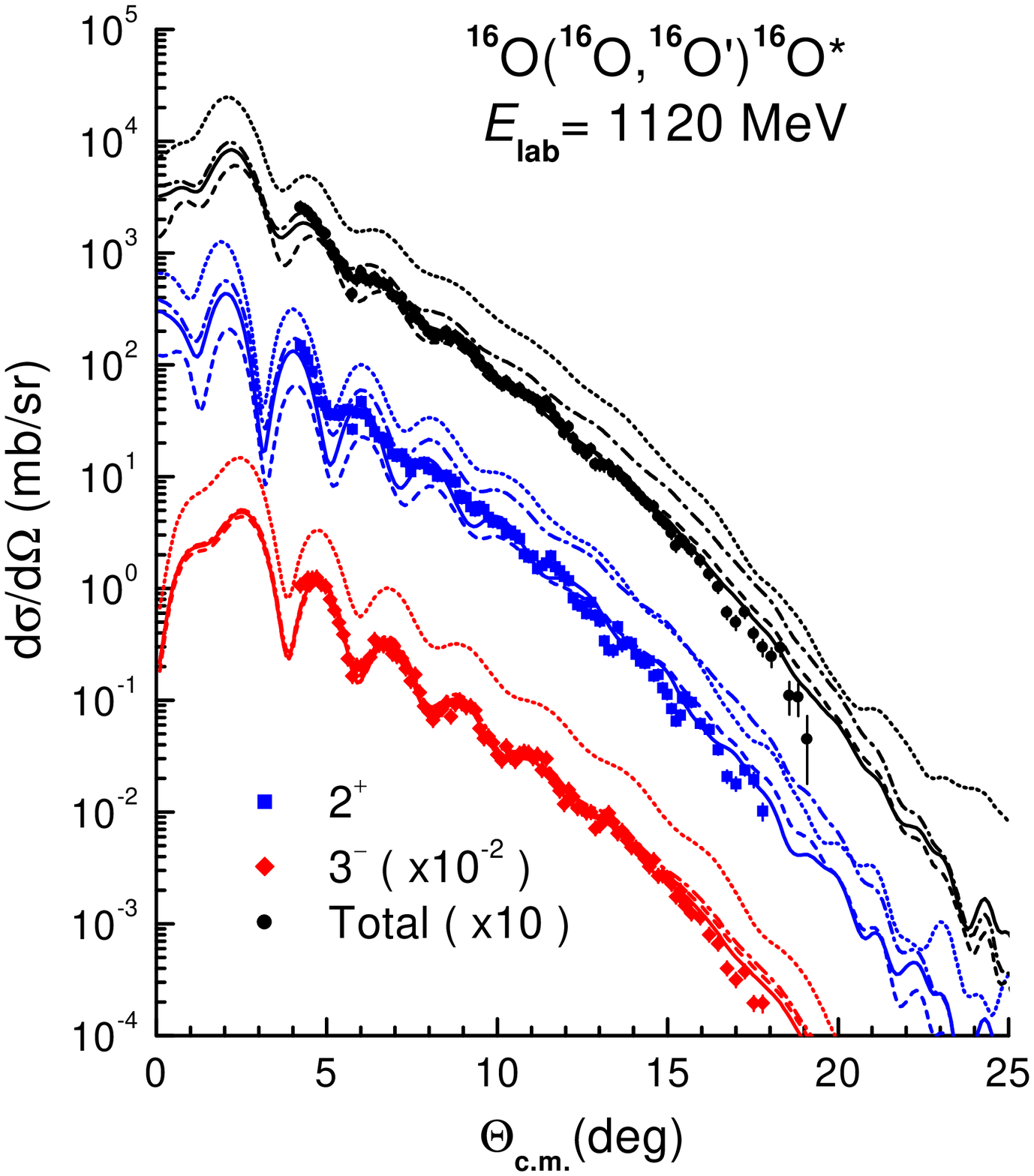}
\caption{The calculated DWBA cross sections for the inelastic \InOO scattering
at $E_{\rm lab}=1120$ MeV in comparison with the measured 2$^+$ and 3$^-$
inelastic cross sections and their sum. The dotted curves are results given by
the original ($N^{(2,3)}_{\rm V}=1$) \emph{complex} inelastic FF [see
Eq.~(\ref{e10})], dash-dotted curves are given by the original \emph{real}
folded FF only, dashed curves are given by the real folded FF which were
renormalized to the best fit the data, and solid curves are obtained with the
original real folded FF but using a \emph{more absorptive} OP in the \ptx exit
channels (see Table~\ref{t1}).} \label{f2}
\end{figure}

As discussed earlier in Ref.~\cite{Kho00}, a very consistent absolute angular
calibration was obtained for the elastic \oo scattering data at $E_{\rm
lab}=1120$ MeV, where all the gauging conditions were well fulfilled for the
target constituents $^6$Li, $^{12}$C, $^{16}$O, $^{40}$Ca and $^{51}$V
simultaneously. The inelastic \InOO scattering data were also quite accurately
measured at this energy, with the inelastic scattering cross sections for the
2$^+$ and 3$^-$ states of $^{16}$O well separated over the whole angular range
covered by the elastic scattering data. Since the DWBA formalism also works
better at medium and high energies (where the coupled-channel effects are not
significant \cite{Sat83}), we started first our DWBA analysis of the inelastic
scattering data at 1120 MeV, and the DWBA results are compared with the measured
\InOO angular distributions in Fig.~\ref{f2}.
Given the deformation lengths fixed above by the measured electric transition
rates of the 2$^+$ and 3$^-$ states and the validity of DWBA, the \emph{complex}
FF's (\ref{e10}) of the inelastic scattering to these states were expected to
give some reasonable description of the data with the unrenormalized folded FF
($N^{(2)}_{\rm V}=N^{(3)}_{\rm V}=1$). However, the DWBA calculation based on
such an assumption, using the same complex OP in the entrance and exit channels,
turned out to grossly overestimate the measured angular distributions for the
2$^+$ and 3$^-$ states of $^{16}$O (see dotted curves in Fig.~\ref{f2}). Even if
we assume lower values for the deformation lengths,
$\delta_{2(3)}\Rightarrow\delta_{2(3)}-\Delta\delta_{2(3)}$ where
$\Delta\delta_{2(3)}$ are the experimental uncertainties of the adopted
$\delta_{2(3)}$ values discussed in Sec.~\ref{sec2}, the absolute 2$^+$ and
3$^-$ cross sections are reduced only by around 10\% and 4\%, respectively. Such
a small change in the inelastic cross section is almost invisible in the
logarithmic scale of the plot.

We tried further to adjust the renormalization factors of the real folded FF's
to best fit the data, using the same imaginary inelastic FF's. These factors
then turned out to be significantly smaller than unity: $N^{(2)}_{\rm V}\approx
0.54$ and $N^{(3)}_{\rm V}\approx 0.06$ which are too unrealistic. Moreover,
with $N^{(3)}_{\rm V}$ approaching almost zero, the strength of the imaginary
3$^-$ inelastic FF is still so strong that the DWBA cross section is roughly
twice the measured cross section over the whole angular range. Keeping
$N^{(3)}_{\rm V}$ value within the range of 0.7-1.0, which is close to
$N^{(0)}_{\rm V}$ values obtained in numerous folding analyses of elastic \aA
and \AA scattering, a reasonable description of the data could only be reached
by switching off the imaginary part of $U^{(3)}_{\rm trans}$ (see dashed and
dash-dotted curves in the lowest part of Fig.~\ref{f2} which were obtained with
$N^{(3)}_{\rm V}=0.938$ and 1, respectively).
\begin{figure}[htb]
\hspace*{-1.5cm} \vspace{-2cm}
\includegraphics[angle=0,scale=0.81]{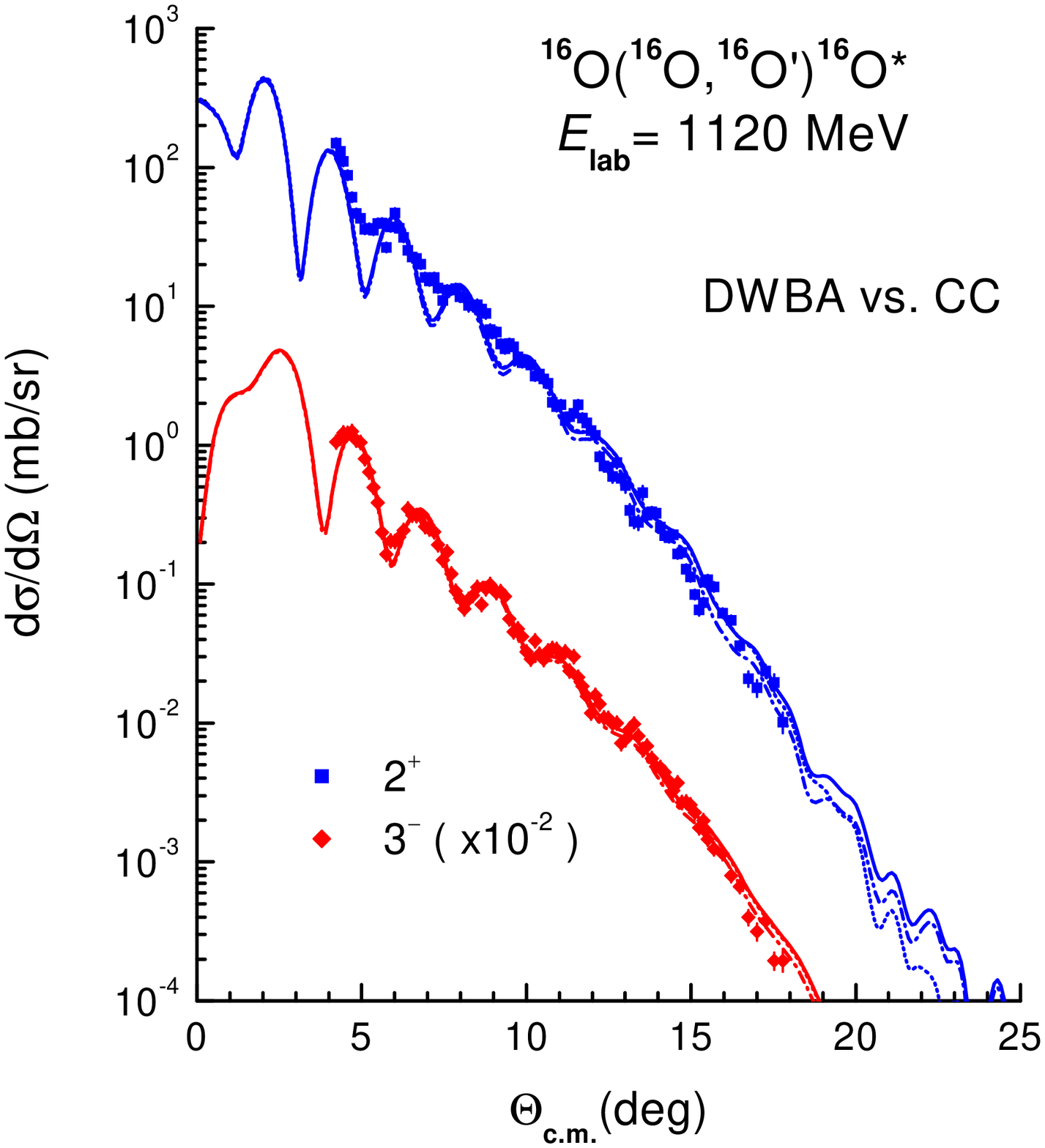}
\caption{The cross sections of the inelastic \InOO scattering at $E_{\rm
lab}=1120$ MeV in comparison with the data. The solid curves were calculated in
the DWBA (see explanation for solid curves in Fig.~\ref{f2}); dash-dotted curves
are given by the CC calculation using the same FF and OP parameters as those
used for solid curves, but taking into account the 3-channel coupling
$2^+\leftrightarrow 0^+\leftrightarrow 3^-$; dotted curves were obtained in the
same CC scheme, but with the WS parameters of the imaginary OP slightly
re-adjusted to best fit the data.} \label{f2cc} \vspace{0cm}
\end{figure}
For the 2$^+$ excitation, even after the imaginary part of $U^{(2)}_{\rm trans}$
is neglected, we could not get a good description of the measured cross section
for the 2$^+$ state over the whole angular range. If the real folded FF is
renormalized by $N^{(2)}_{\rm V}\approx 0.63$ (a strong \emph{quenching} of the
transition strength), the DWBA result fits reasonably the data at large angles
but underestimates significantly the 2$^+$ data at forward angles (dashed curve
in the middle part of Fig.~\ref{f2}). When the real folded FF is kept unchanged
($N^{(2,3)}_{\rm V}=1$), the DWBA calculation can only fit the data points at
smallest angles and strongly overestimate the data at large angles (dash-dotted
curve in the middle part of Fig.~\ref{f2}). An overall good description of the
measured 2$^+$ cross section could finally be reached only when a more
absorptive OP is used for the \2x exit channel, with the absorption strength
around 40 to 60\% stronger than that in the entrance channel (see
Table~\ref{t1}). In this case, we have used the real folded FF with
$N^{(2,3)}_{\rm V}=1$ and the same WS geometry of the \ptx imaginary OP as that
in the \oo entrance channel, only the WS strengths of the imaginary OP in the
exit channel were adjusted to obtain a good description of the inelastic
scattering data by the DWBA calculation. For the 3$^-$ excitation, such a
procedure lead only to a slightly more absorptive OP in the \3x exit. The
situation remains nearly the same when we switch on the 3-channel coupling
$2^+\leftrightarrow 0^+\leftrightarrow 3^-$. The coupled-channel (CC)
calculation using the same OP and inelastic FF as those used in the DWBA
calculation leads to a slight reduction (around 10-30\%) of the inelastic cross
sections at largest angles (compare solid and dash-dotted curves in
Fig.~\ref{f2cc}). After the depths of the WS imaginary potentials were
re-adjusted to fit the data in the CC calculation (the dotted curves in
Fig.~\ref{f2cc}), the absorptive strength of the OP is reduced by only around
5-6\%. Such a weak CC effect shown in Fig.~\ref{f2cc} confirms the conclusion
made based on the DWBA results plotted in Fig.~\ref{f2}.

\begin{figure}[htb]
\hspace*{-1cm} \vspace{-1.5cm}
\includegraphics[angle=0,scale=0.81]{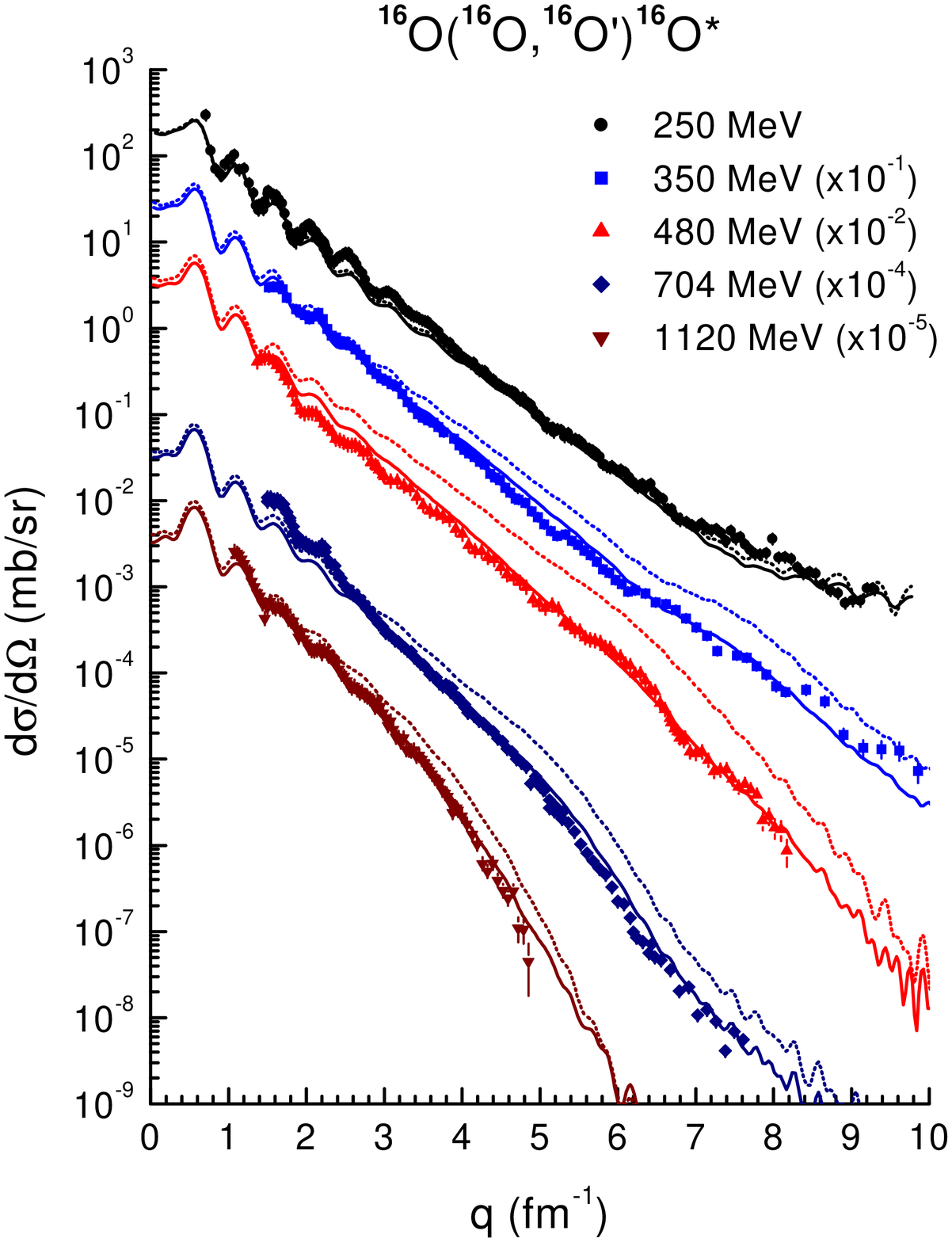}
\caption{The DWBA description of the measured total (2$^++3^-$) inelastic \InOO
scattering cross sections at $E_{\rm lab}=250-1120$ MeV given by the original
($N^{(2,3)}_{\rm V}=1$) real folded FF. The dotted curves were obtained with the
same complex OP for the entrance and exit channels, and the solid curves were
obtained with a more absorptive OP in the exit channels (see Table~\ref{t1}).
The cross sections are plotted versus the momentum transfer
$q=2k\sin(\Theta_{\rm c.m.}/2)$, where $k$ is the wave number of the
projectile.} \label{f3} \vspace{0cm}
\end{figure}
We have further analyzed the \InOO data at the lower energies. Since the
excitation energies of the 2$^+$ and 3$^-$ states of $^{16}$O are separated by
about 0.8 MeV, the inelastic cross sections for these two states could not be
well separated at large angles due to stronger kinematical effects. There only
the sum of the 2$^+$ and 3$^-$ cross sections could be measured. We had to
maintain, therefore, in each case a good DWBA description of both the 2$^+$ and
3$^-$ data at forward angles as well as their sum at large angles, in the
refractive angular region. Our DWBA analysis of the considered inelastic \InOO
data at lower energies shows consistently that the inelastic scattering to the
lowest 2$^+$ and 3$^-$ states of $^{16}$O exhausts mainly the strength of the
\emph{real} part of the inelastic FF (\ref{e10}).

If we consider the imaginary part of the inelastic FF as originating from a
`dynamic polarization' of the transition potential, similar to that of the
microscopic OP according to Feshbach \cite{Fes92}, then these results would
indicate a dominance of the direct (one-step) inelastic scattering process and
the contribution from higher order terms to the \InOO reaction is negligible. A
similar conclusion has also been made recently by Katsuma {\sl et al.}
\cite{Kat02} based on their detailed CC analysis of the elastic and inelastic
\oo scattering at 350 MeV.

\begin{figure}[htb]
\hspace*{-2cm} \vspace{0cm}
\includegraphics[angle=270,scale=0.65]{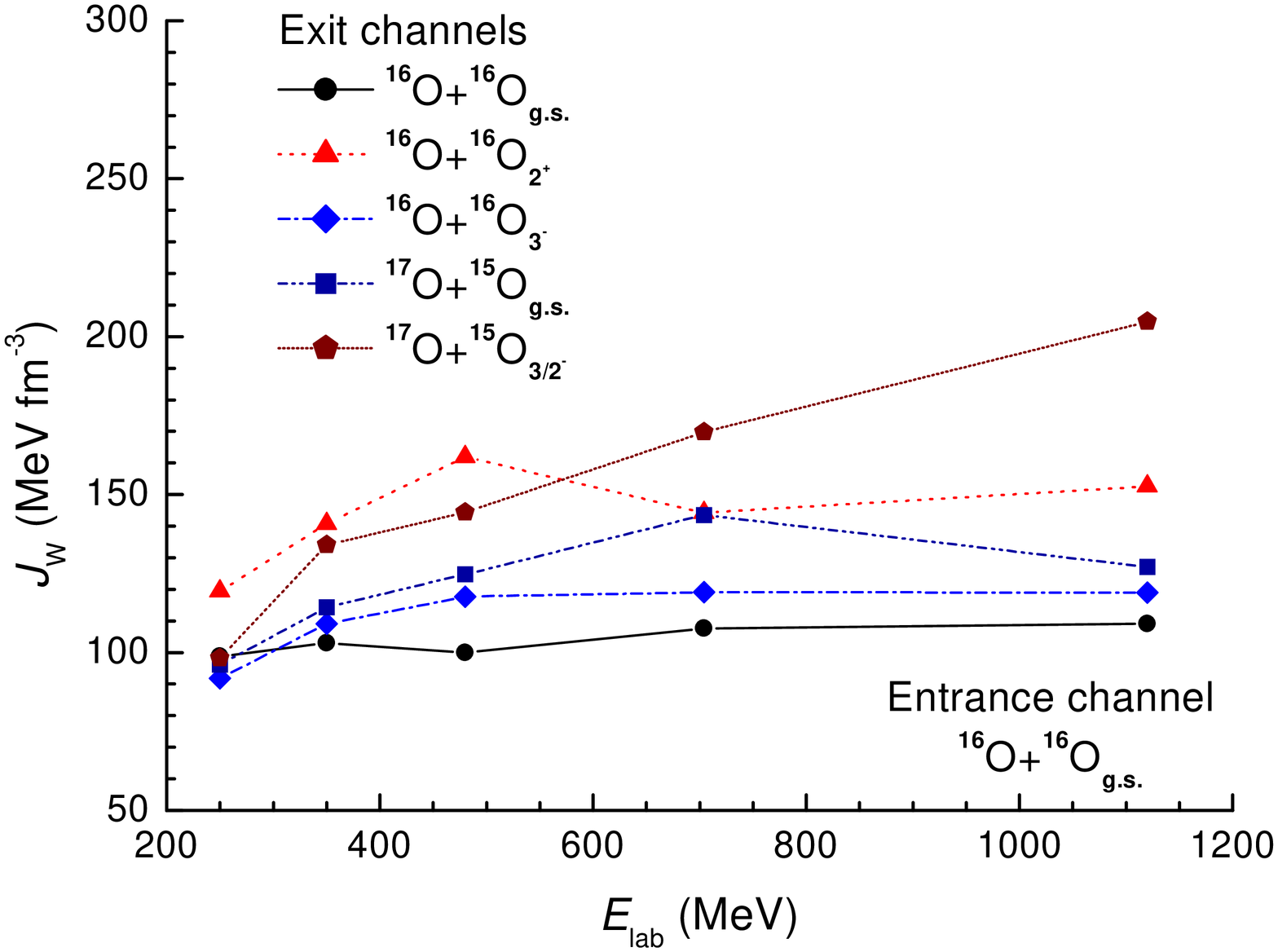}
\caption{Energy dependence of the volume integral per interacting nucleon pair
$J_{\rm W}$ of the WS imaginary OP used for the elastic \oo channel, \2x and \3x
exit channels of the inelastic \InOO scattering, and the exit channels of the
one-neutron \o17o15 transfer \cite{Boh02} to the ground state $^{15}$O$_{\rm
g.s.}$ and the excited state $^{15}$O$_{{3/2}^-}$. The lines are only to guide
the eye.} \label{f3j} \vspace{1cm}
\end{figure}

The DWBA results obtained with the inelastic FF consisting only of the real
folded FF are shown in Fig.~\ref{f3}. Since a renormalization of the real folded
FF usually could not deliver a good DWBA description of the data over the whole
angular range and given the dominance of the direct one-step scattering process,
we have used throughout our DWBA analysis $N^{(\lambda)}_{\rm V}=1$ and
$\delta_{\lambda}$ values given in Sec.~\ref{sec2}. These DWBA results confirm
again that a significantly more absorptive OP must be used for the exit
channels, especially for \2x where the absorption was found roughly 40-50\%
stronger than that in the \oo entrance channel (see the volume integral $J_{\rm
W}$ of the absorptive WS potential in Table~\ref{t1}). We note that the
enhancement of the absorption found for the \ptx exit channels is very similar
to the enhanced absorption established earlier in our (finite-range) DWBA
analysis of the one-neutron transfer \o17o15 reaction \cite{Boh02}, where the
absorption in the $^{17}$O+$^{15}$O exit channels is also about 50\% or more
stronger than that in the \oo entrance, with the corresponding volume integrals
$J_{\rm W}$ of the absorptive WS potential close to those normally observed for
a HI system with strong absorption. Since the enhanced absorption implies a
shorter mean free path \cite{Jeu76,Neg81}, our DWBA results for the inelastic
\InOO scattering also support the conclusion of Ref.~\cite{Boh02} that the
nuclear mean free path is quite different as one considers different reaction
channels. The nuclear mean free path in the overlapping region was shown
\cite{Boh02} to gradually decrease when one goes from a tightly bound and
spherical double closed-shell nucleus $^{16}$O to a less bound $^{15}$O nucleus
(in its ground and excited p$_{3/2}$ states). Now we see the same trend for the
$^{16}$O nucleus in the excited 2$^+$ and 3$^-$ states. Thus, the enhanced
absorption found in the exit channels should, in general, imply a shorter mean
free path of the excited (and less-bound) nuclear cluster in the nuclear medium.
To illustrate the trend of the increased absorption in the exit channels of the
quasi-elastic \oo scattering, we have plotted in Fig.~\ref{f3j} the volume
integral of the absorptive WS potential $J_{\rm W}$ per interacting nucleon
pair. While a volume integral $J_{\rm W}\approx 100$ MeV~fm$^3$ was found for
the elastic \oo scattering channel, with the weakening or even disappearance of
refractive effects in the inelastic \InOO scattering and one-neutron \o17o15
transfer channels \cite{Boh02} the correct description of the data over a large
angular range required consistently an increased absorption in the exit
channels. For the inelastic scattering to the excited 2$^+$ state of $^{16}$O
and one-neutron transfer to the excited $3/2^-$ state of $^{15}$O, the enhanced
absorption has lead to a volume integral $J_{\rm W}$ up to around 150-200
MeV~fm$^3$  (see Fig.~\ref{f3j}) which is close to that observed for
strong-absorbing HI systems. For example, the OM analysis of the elastic
scattering of 390 MeV $^{20}$Ne on $^{12}$C using the real WS and folded
potentials has given $J_{\rm W}=207$ MeV~fm$^3$ and 155 MeV~fm$^3$, respectively
\cite{Boh93}.

The enhanced absorption found for the exit channels of the inelastic \oo
scattering and \o17o15 transfer reaction stresses the need to have a realistic
choice for the OP not only in the entrance but also in the exit channel. The use
of the same complex OP in both the entrance and exit channels might lead to a
large uncertainty in the deduced transition strength if one follows the standard
method of scaling the inelastic FF to match the DWBA results to the measured
angular distributions. In particular, one needs to be aware of this effect while
analyzing the inelastic scattering data measured at `refractive' energies for a
light HI system containing an unstable nucleus with the excited state being
either unbound or very weakly bound. We note finally that the subtle effect of
absorption enhancement could have been found only owing to very accurately
measured inelastic \InOO scattering and one-neutron \o17o15 transfer data which
cover a wide angular range and about 6 orders of the cross-section magnitude.

\begin{figure}[htb]
\hspace*{-1.5cm} \vspace{-2cm}
\includegraphics[angle=0,scale=0.81]{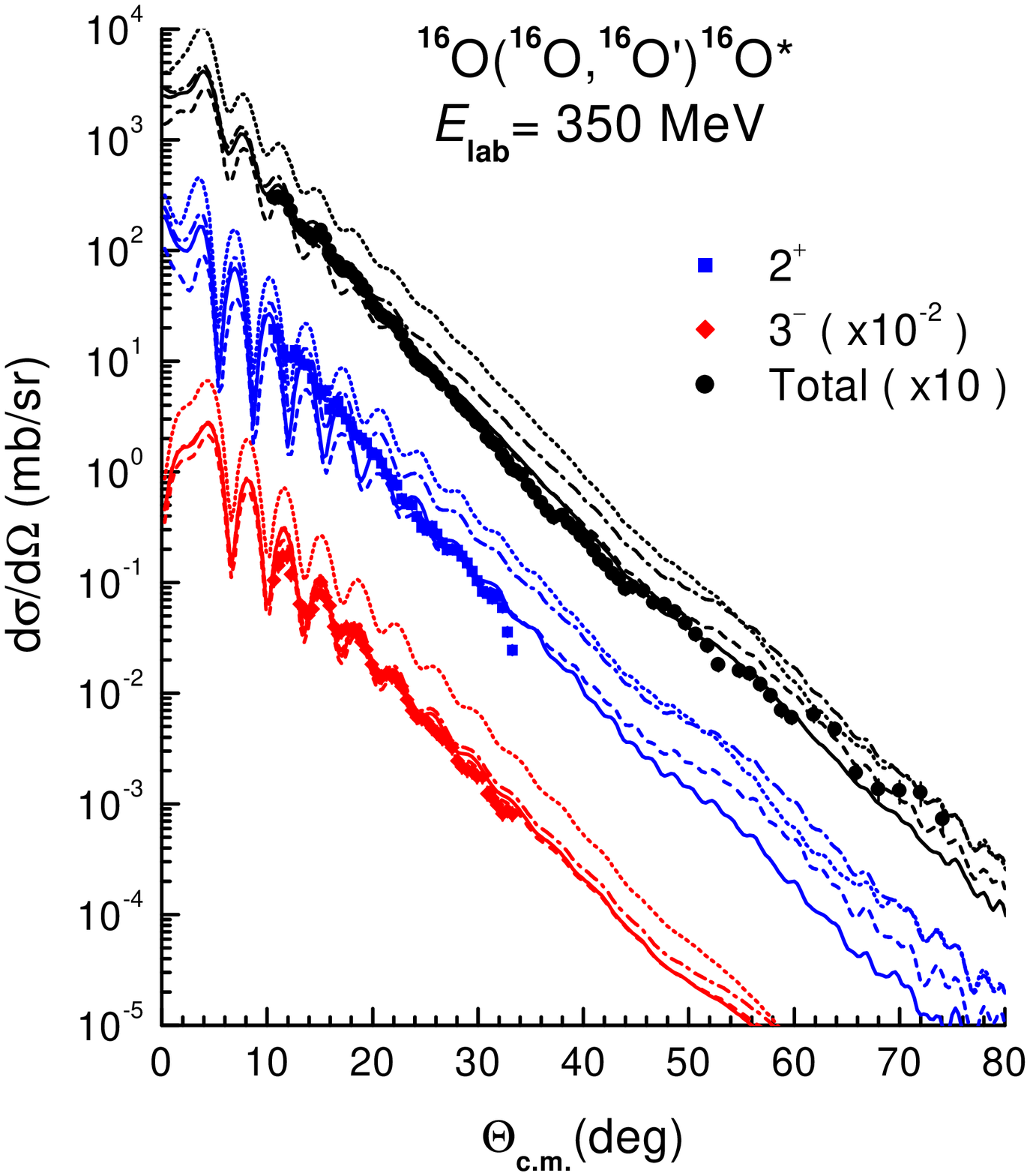}
\caption{The same as Fig.~\ref{f2} but for $E_{\rm lab}=350$ MeV.} \label{f4}
\vspace{1cm}
\end{figure}
We consider now the refractive features in the inelastic \InOO scattering in
more details, and the best data set for this purpose are those measured at
$E_{\rm lab}=350$ MeV. At this energy, a broad shoulder-like primary rainbow
pattern preceded by the first Airy minimum at $\Theta_{\rm c.m.}\approx
44^\circ$ (or $q\approx 6.1$ fm$^{-1}$) has been observed in the elastic \oo
scattering cross section (see Fig.~\ref{f1}). From the DWBA results given by the
same OP in the entrance and exit channel (dotted and dash-dotted curves) for the
inelastic scattering at 350 MeV shown in Fig.~\ref{f4} one can see a weak rise
in the sum of the 2$^+$  and $3^-$ cross sections at $\Theta_{\rm c.m.}\geq
46^\circ$. This broad pattern at large angles is about the same for the 2$^+$
and $3^-$ cross sections (with overlayed oscillatory structures at the forward
angles which are out of phase) and should be of the refractive nature and caused
by the same interference mechanism that gives rise to the Airy oscillation seen
in the elastic \oo scattering \cite{Mic04}. Note that the rainbow shoulder is
slightly shifted towards large angles because of a small decrease of the c.m.
energy in the exit channel. This structure becomes weaker if one uses a more
absorptive OP in the exit channels to reproduce the measured (2$^++3^-$) cross
section as discussed above. Nevertheless, a remnant of the primary rainbow
shoulder can still be seen separately in the calculated 2$^+$ and $3^-$ cross
sections as well as in the measured total cross section. From the calculated and
measured \InOO cross sections at all energies shown in Fig.~\ref{f3} one can
trace this weak rainbow pattern in the inelastic scattering cross section from
the energy of 350 MeV up to 704 MeV.

\begin{figure}[htb]
\hspace*{-1.5cm} \vspace{-2cm}
\includegraphics[angle=0,scale=0.81]{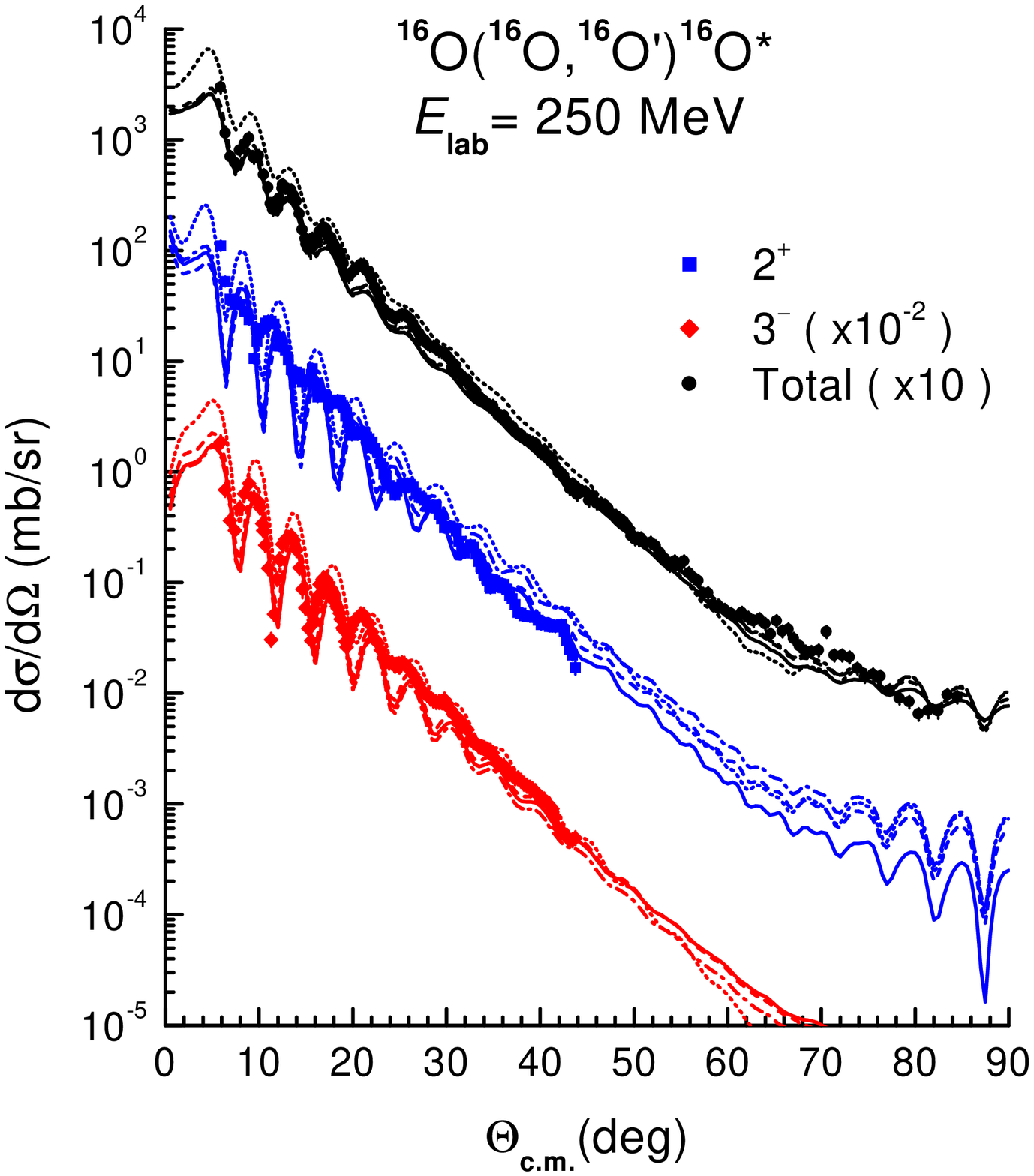}
\caption{The same as Fig.~\ref{f2} but for $E_{\rm lab}=250$ MeV.} \label{f5}
\vspace{1cm}
\end{figure}
At 250 MeV, the nuclear rainbow pattern of the elastic \oo scattering was shown
to be shifted to scattering angles larger than $90^\circ$ \cite{Kho00} and there
it is invisible due to the strong Mott interference. That is also the reason why
one can not see it in the inelastic \InOO scattering. Interestingly, the
enhancement of absorption found at higher energies is not very obvious at 250
MeV where the results given by the same OP for both the entrance and exit
channels deliver also a fairly good description of the measured data (see
Fig.~\ref{f5}). We have seen the same trend in our recent DWBA analysis of the
one-neutron \o17o15 reaction \cite{Boh02}, where the enhanced absorption in the
exit channel was also not observed at 250 MeV. It is very likely that the
enhancement of the absorption in the exit channel is a feature associated with
the refractive nature of quasi-elastic \AA scattering. Such an effect can be
well expected from the interpretation of the rainbow scattering as a phenomenon
associated with a \emph{weak absorption} which allows a \emph{deeper
interpenetration} of the two nuclei, with the refractive pattern determined by
the \AA potential at small distances \cite{vOe03}.

\begin{figure}[htb]
\hspace*{-1cm} \vspace{-2cm}
\includegraphics[angle=0,scale=0.77]{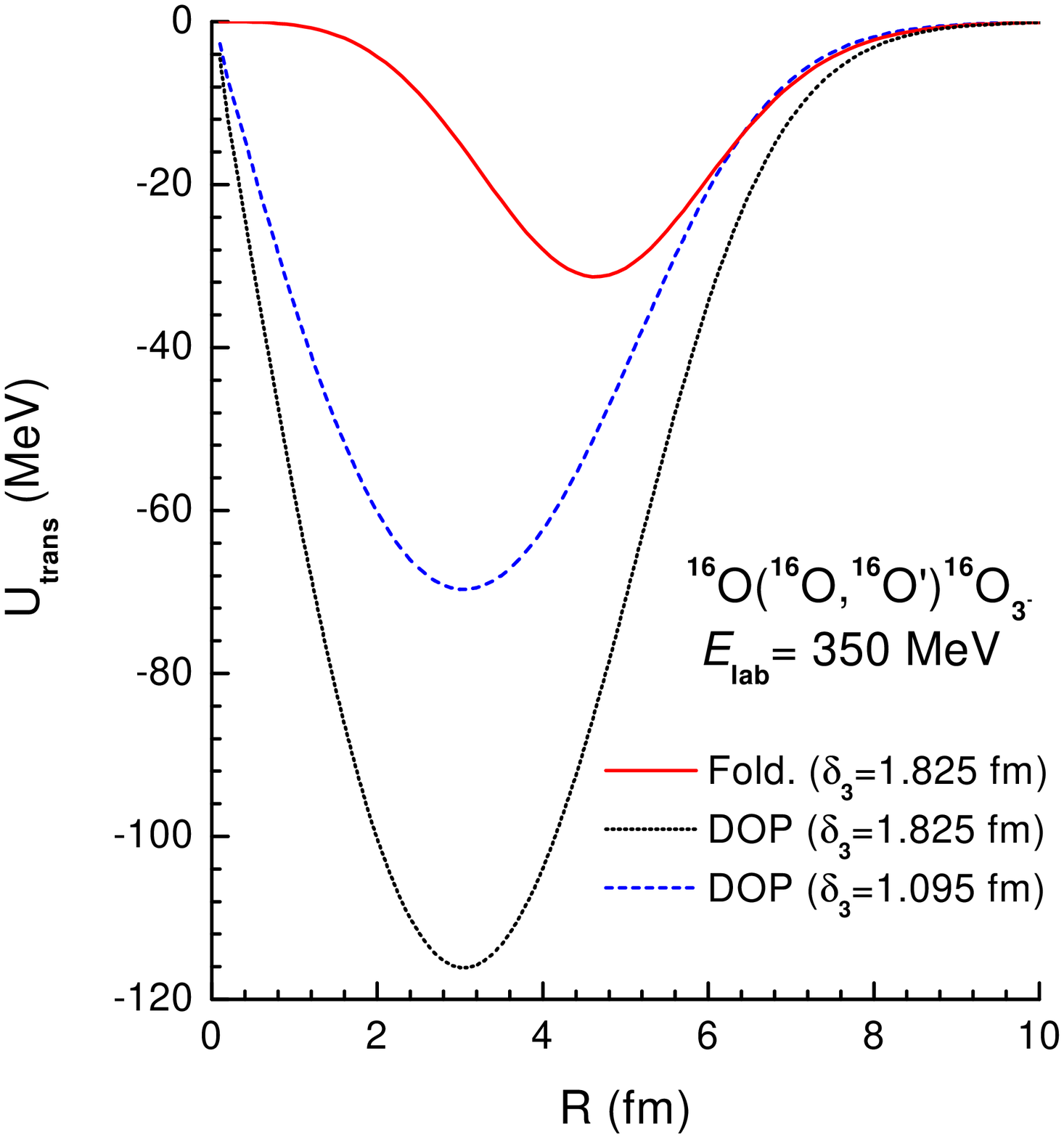}
\caption{The form factors for the transition $0^+\rightarrow 3^-$ in the
inelastic \oo scattering at 350 MeV. The folded $V^{(3)}_{\rm F}$ form factor is
the solid curve, the DOP curves represent the best-fit elastic folded potential
deformed with the same deformation length $\delta_3$ as that used to generate
$V^{(3)}_{\rm F}$ (dotted curve) or with $\delta_3$ adjusted to the best DWBA
fit to the inelastic scattering data (dashed curve).} \label{f6} \vspace{1cm}
\end{figure}

\begin{figure}[htb]
\hspace*{-1.5cm} \vspace{-2cm}
\includegraphics[angle=0,scale=0.81]{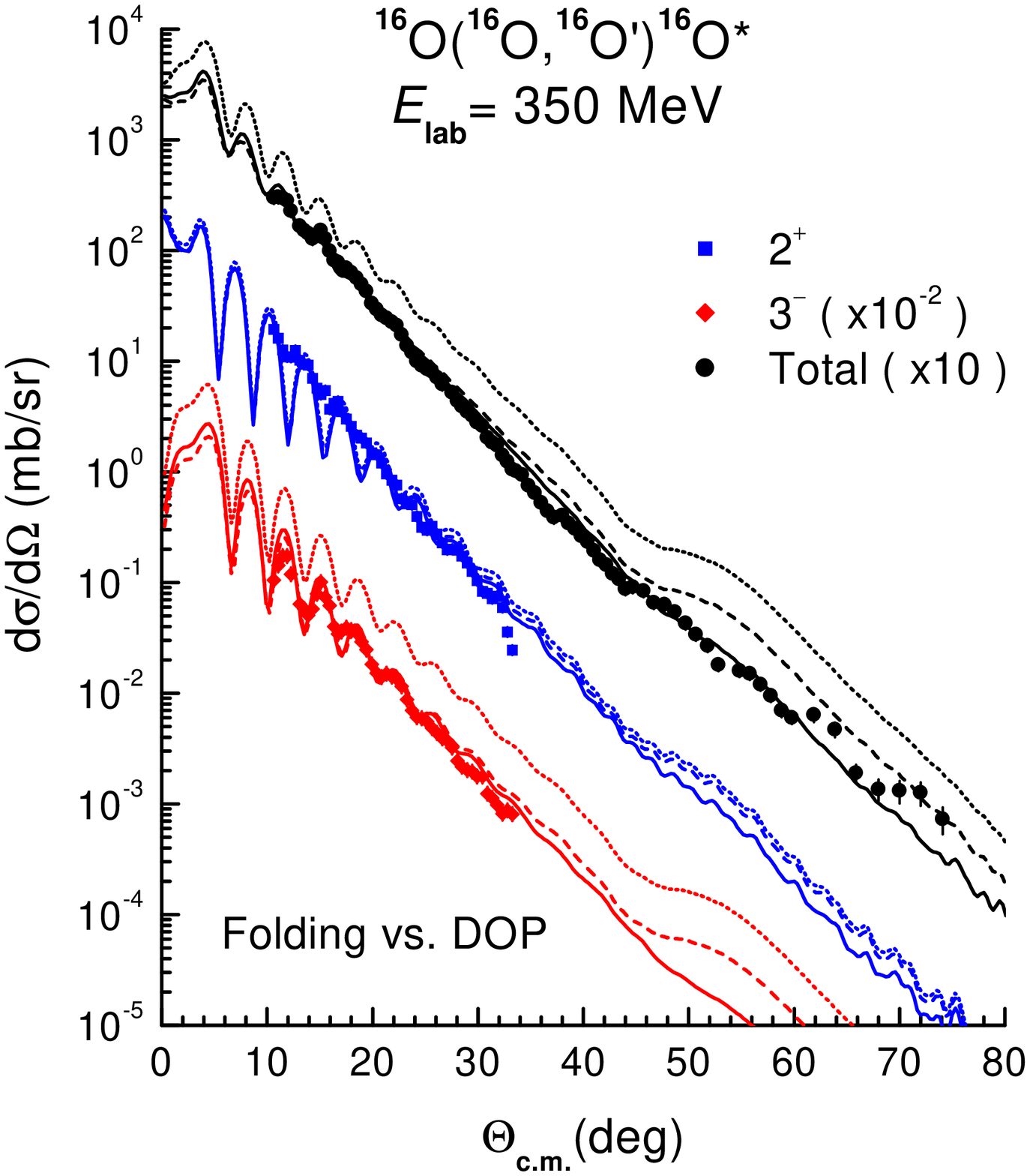}
\caption{The calculated DWBA cross sections for the inelastic \InOO scattering
at $E_{\rm lab}=350$ MeV in comparison with the measured 2$^+$ and 3$^-$
inelastic cross sections and their sum. The solid curves were obtained with the
same folded form factors $V^{(2,3)}_{\rm F}$ as those discussed in
Fig.~\ref{f4}, dotted curves were obtained with the DOP form factors built upon
the same deformation lengths $\delta_{2,3}$ as those used to generate
$V^{(2,3)}_{\rm F}$, and dashed curves are the DOP results but with
$\delta_{2,3}$ adjusted to the best DWBA fit to the data.} \label{f7}
\vspace{1cm}
\end{figure}

Given a weak rainbow pattern seen in the inelastic scattering cross section at
the energies of 350-704 MeV, one might also expect some sensibility of the
measured inelastic data to the shape of the \AA form factor at small distances.
To probe this effect, we have constructed the transition form factors
(\ref{e2}) using both the folding and DOP methods. To obtain the DOP form
factors, we have first deformed the (best-fit) elastic folded potential by the
same deformation lengths $\delta_\lambda$ as that used to generate the nuclear
transition densities (\ref{e8}) for the folding calculation. Although the two
choices of the FF have more or less the same strength at the surface of the
dinuclear system, they differ from each other substantially at small radii
(compare the solid and dotted curves in Fig.~\ref{f6}). This difference shows
up clearly in the calculated inelastic cross section and one can see from
Fig.~\ref{f7} that a reasonable DWBA description of the data by the DOP form
factors can only be reached when the deformation lengths $\delta_{2,3}$ used in
the DOP model are substantially reduced, especially, for the $\lambda=3$ case
where $\delta_3$ has to be reduced by a factor around 1.67. The corresponding
transition rate is then reduced to $B(E3\uparrow)\approx 531\ e^2$fm$^6$ which
is nearly three times smaller than the adopted experimental value of $1480\pm
50\ e^2$fm$^6$ \cite{Kib02}. Such a deficiency of the DOP form factor, which
leads to an artificial ``hindrance" of the $3^-$ excitation strength, has been
first pointed out by Beene et al. \cite{Be95} and is now confirmed again by our
folding + DWBA analysis of the inelastic \InOO scattering. Moreover, with the
inelastic scattering data covering a wide angular range, it can be seen in
Fig.~\ref{f7} that the DOP method completely fails to give a good description
of the data points at largest angles which are known to be sensitive to the
shape of the FF at small distances. Like the rainbow pattern in the elastic \oo
scattering which can be used to probe the real OP at small internuclear
distances \cite{vOe03}, we have established that the weak remnant of the
primary rainbow observed in the inelastic \InOO scattering at 350 MeV is also
helpful in testing the inelastic FF at small radii.

\section{Summary}
 \label{sec4}
Inelastic \InOO scattering data at $E_{\rm lab}=250, 350, 480, 704$ and 1120
MeV for the transitions to the lowest 2$^+$ and $3^-$ states in $^{16}$O,
covering the same wide angular range as that covered by the elastic \oo
scattering at these energies \cite{Boh93,Sti89,Bar96,Nuo98}, have been measured
and analyzed within the DWBA using the semi-microscopic optical potential and
inelastic form factor given by the folding model \cite{KhoSat}. Although the
refractive pattern of the inelastic \InOO scattering was found less pronounced
compared to that observed in the elastic scattering channel, a similar
evolution of the primary rainbow remnant could still be traced in the inelastic
scattering cross section up to $E_{\rm lab}=704$ MeV.

Given the strengths of the folded form factors fixed by the deformation lengths
determined from the experimental $B(E2\uparrow)$ and $B(E3\uparrow)$ data
\cite{Ram01,Kib02}, a reasonable DWBA description of the measured 2$^+$ and
$3^-$ angular distributions has been obtained only if the absorption in exit
channels is significantly increased, especially for the
$^{16}$O+$^{16}$O$_{2^+}$ exit, at the energies where the refractive (rainbow)
pattern was well observed in the elastic scattering channel. Our DWBA analysis
also shows consistently that the considered inelastic scattering to the lowest
2$^+$ and 3$^-$ states of $^{16}$O exhausts mainly the strength of the
\emph{real} part of the inelastic form factor (\ref{e10}). In terms of
Feshbach's formalism \cite{Fes92}, these results should indicate the dominance
of the direct (one-step) inelastic scattering and negligible contribution from
higher order terms to the \InOO reaction.

The enhanced absorption found for the exit channels of the inelastic \oo
scattering and \o17o15 transfer reaction \cite{Boh02} indicates that the
nuclear mean free path is decreasing during the transition from the entrance
channel containing two tightly bound double closed-shell $^{16}$O nuclei in
their ground states to a \emph{less bound} exit channel containing either
$^{15}$O nucleus or $^{16}$O nucleus in the excited 2$^+$ or 3$^-$ states. This
would be of interest to study this effect in the future analysis of the
quasi-elastic scattering data measured for the \AA systems involving unstable
nuclei, where the partitions in the entrance and exit channels are very
differently bound.

The weak rainbow remnant observed in the inelastic \InOO scattering was shown
to be quite helpful in testing the inelastic FF at small radii. A comparison
made between the folding and DOP models for the inelastic form factor shows
clearly the failure of the DOP approach in the description of the inelastic
\InOO scattering data at 350 MeV, which cover both the diffractive and
refractive regions in the angular distribution. The use of the DOP model also
significantly underestimates the deformation lengths of the nuclear
excitations, especially for $\lambda\geq 3$. Therefore, we recommend again the
folding model as a more reliable tool in the analysis of the inelastic \AA
scattering.

\section*{Acknowledgements}
One of the authors (D.T.K.) thanks the Hahn-Meitner-Institut Berlin and the
Alexander-von-Humboldt Stiftung of Germany for the financial support and
hospitality during his stays at HMI Berlin in 2000, 2001 and 2003 to work on
this project. The comment by Ken Amos on the kinematical transformation of the
NN interaction from the NN frame to the \AA frame is also appreciated.

\end{document}